\documentclass[12pt,preprint]{aastex}
\slugcomment{Submitted to ApJL 02/03/04 05:00 PM}

\begin{document}

\title{High-Resolution Measurements Of Intersystem Bands Of Carbon Monoxide
Toward X Persei}
\author{{\sc Yaron Sheffer\altaffilmark{1}, S. R. Federman\altaffilmark{1},
and David L. Lambert\altaffilmark{2}}}

\altaffiltext{1}{Department of Physics and Astronomy, University of Toledo, Toledo, OH 43606;
ysheffer@physics.utoledo.edu, sfederm@uoft02.utoledo.edu.}
\altaffiltext{2}{Department of Astronomy, University of Texas, Austin, TX 78712;
dll@astro.as.utexas.edu.}

\begin{abstract}
In an echelle spectrum of \objectname{X Per} acquired with the Space Telescope Imaging Spectrograph
we have identified individual rotational lines of 11 triplet-singlet (intersystem) absorption bands
of $^{12}$CO. Four bands provide first detections for interstellar clouds. From a comparison
with the $\zeta$ Oph sight line we find that X Per is obscured by a higher $^{12}$CO column density
of 1.4~$\times$ 10$^{16}$ cm$^{-2}$. Together with the high spectral resolution of
1.3 km~s$^{-1}$, this allows (i) an improved measurement of previously published $f$-values for
seven bands, and (ii) an extraction of the first astrophysical oscillator strengths for
{\it d$-$X} (8$-$0), (9$-$0), and (10$-$0), as well as for {\it e$-$X} (12$-$0).
The $^{13}$CO {\it d$-$X} (12$-$0) band, previously suspected to exist 
toward $\zeta$ Oph, is now readily resolved and modeled. Our derived intersystem $f$-values for
$^{12}$CO include a few mild ($\leq$ 34\%) disagreements with recent predictions from a
perturbation analysis calculated for the interstellar excitation temperature. Overall, the
comparison confirms the superiority of employing multiple singlet levels in the calculations
of mixing coefficients over previous single-level predictions.
\end{abstract}


\section{INTRODUCTION}

Observers of interstellar sight lines are familiar with the strong
$A~^1\Pi$$-$$X~^1\Sigma^+$ permitted absorption bands of CO below 1545 \AA.
Less familiar to observers are the intersystem bands of CO, which involve the triplet states
$a^\prime~^3\Sigma^+$, $d~^3\Delta_i$, and $e~^3\Sigma^-$,
whose electric dipole transitions to the ground state $X~^1\Sigma^+$
are spin-forbidden. These triplet states have potential energy curves that cross the curve
of $A~^1\Pi$---see Figure 1 of Rostas et al. (2000)---and via mutual perturbations acquire a small
percentage of $^1\Pi$ character at the expense of the singlet state. The resulting
triplet-singlet $f$-values are relatively small, thus turning the intersystem bands into very
useful astrophysical probes along interstellar (IS) sight lines for which the {\it A$-$X} bands
are optically thick, see \citet[hereafter MN]{mn94}, \citet[hereafter F94]{f94}.
To exploit such probes, accurate $f$-values of intersystem transitions are required.
\citet[hereafter Rostas et al.]{r00} presented predictions from
extended calculations of the triplet-singlet mixing coefficients,
partly as an attempt to resolve discrepancies between previous theoretical (MN) and
astrophysical $f$-values. Rostas et al. also conducted low-resolution laboratory measurements
at room temperature to improve upon the experimental results of \citet{e92}. High-resolution
laboratory measurements on two intersystem bands were also reported by \citet{stark}.

Using the {\it Hubble Space Telescope\/}\footnotemark[3]
\footnotetext[3]{Based on observations obtained with the NASA/ESA {\it Hubble Space Telescope}
through the Space Telescope Science Institute, which is operated by the Association of Universities
for Research in Astronomy, Inc., under NASA contract NAS5-26555.}
($HST$), we obtained a high-resolution echelle spectrum
of the O9.5V star X Per (HD 24534) with the Space Telescope Imaging Spectrograph
(STIS). Permitted {\it A$-$X} bands along this line of sight were first modeled by \citet{ka}, who
based his analysis of $^{12}$CO on archival $HST$ Goddard High-Resolution Spectrograph (GHRS)
observations. For our STIS observations, the weakest permitted band available for analysis
is the {\it A$-$X} (8$-$0) band, which still has a large optical depth ($\tau$)
in the cores of the strongest lines, $\tau$($R$(0))~$\sim$ 70. 
Therefore, the column density ($N$) of $^{12}$CO toward X Per can be measured more reliably
from the observed intersystem bands, which have a much smaller $\tau$.
In this Letter, we determine $N$($^{12}$CO) through a comparison with our previous
observations of intersystem bands toward $\zeta$ Oph (F94). Then we derive IS $f$-values toward
X Per for all seven ``old'' bands and for five intersystem bands seen in IS spectra for the first
time. Finally, we compare our $f$-values with the predictions of Rostas et al.
Regarding nomenclature, we make use in this paper of the short notation introduced by
Rostas et al., where, e.g., $d$12 stands for {\it d$-$X} (12$-$0).

\section{OBSERVATIONS AND MODELING}

STIS spectra of X Per were acquired in 2001 February and March with grating E140H
during 10,728~s over five orbits, yielding the data sets o64812010$-$030 and o64813010$-$020.
The star was observed through
the smallest aperture (0.1\arcsec~$\times$ 0.03\arcsec), which was designed to provide
the highest resolving power, $R$~= 200,000 \citep{jt01}.
Elaborate reduction procedures had to be employed in order to correct for odd-even
sub-pixel effects in the unbinned (2048-pixel) exposures. Normal reduction routines were then
followed by processing the spectra in the STSDAS environment of IRAF.
The five orbital exposures were co-added into a single spectrum of 42 orders,
sporting a signal-to-noise ratio of 70, or 100
for features detected and combined from two adjacent overlapping orders.
Since the continuum is modulated by shallow photospheric features from X Per,
short spectral segments centered on features (bands) of interest were rectified
and normalized before being compared to a fitted synthetic model of the pertinent feature.
Flux and wavelength calibrations were performed by the
Space Telescope Science Institute pipeline and supplied with the binned, 1024-pixel
exposures.

A previous analysis of $^{12}$CO by \citet{kb} inferred that this line of sight
involves two unresolved velocity components.  However, the significantly higher resolution
of our spectrum shows profile asymmetries due to underlying structure, which
required four components for a good fit. This is the same number of components
found in preliminary fits of high-resolution \ion{K}{1} spectra of X Per (D. E. Welty, private
communication), although we derived a different set of relative velocities, fractions, and
component widths. Some 87\% of the total $N$($^{12}$CO) resides in a main component with
a Doppler parameter ($b$-value) of 0.39 km~s$^{-1}$.
The other three components have between 2\% and 9\% shares of $N$, with
$b$-values between 0.17 and 2.09 km~s$^{-1}$, as will be fully described in a future paper.
The highly saturated {\it A$-$X} bands could be fitted only once the total $N$
was established via the intersystem bands.
Our method of analysis employed a multi-parametric spectrum synthesis code based on
Voigt profile line transfer equations given in \citet{bvd}.
The computed absorption spectrum was convolved with a Gaussian instrumental profile
(Jenkins \& Tripp 2001) and then matched with the data via
rms-minimization down to relative parameter steps of 10$^{-4}$ or less.

\section{THE COLUMN DENSITY OF $^{12}$CO}

In order to determine $f$-values for intersystem bands, $N$($^{12}$CO) toward X Per
must be known. For optically thin features, the equivalent width ($W_{\lambda}$) is proportional
to $\tau$, i.e., $W_{\lambda}$~$\propto$ $fN$. We can, therefore,
utilize the published value of $N$ toward $\zeta$ Oph
and the ratio of $W_{\lambda}$s for optically thin bands toward the two stars
in the determination of $N$ toward X Per.
The result should be independent of the intersystem $f$-values
and of the modeled cloud structure, provided we exclude the $W_{\lambda}$
ratio for the optically thick $a^\prime$14.
Column 4 of Table 1 lists the ratios of band $W_{\lambda}$ toward X Per
over band $W_{\lambda}$ toward $\zeta$ Oph.
Among the optically thin bands, the highest $W_{\lambda}$ ratio belongs to $e$4,
which was the faintest intersystem band analyzed by F94.
In addition, $e$4 also sports the largest
deviation of derived $f$-value in the low-$\tau$ sample of F94.
It is clear that F94 underestimated the low-resolution $W_{\lambda}$ for $e$4
due to placing the continuum too low (see their Figure 1).
Dropping the $e$4 band, and thus avoiding a 5\% increase in the average,
we obtain $W_\lambda$(X~Per)/$W_\lambda$($\zeta$~Oph)~= 5.54~$\pm$ 0.87.
Under the necessary and fulfilled condition of optical thinness, this is also
the ratio of $N$(X~Per) over $N$($\zeta$~Oph). Here we shall adopt the \citet{l94} column density
toward $\zeta$ Oph, $N$($^{12}$CO)~= (2.54~$\pm$ 0.16)~$\times$ 10$^{15}$ cm$^{-2}$,
which was also employed by F94. Based on the measured $N$ ratio, we infer that
$N$($^{12}$CO) toward X Per is (1.41~$\pm$ 0.22)~$\times$ 10$^{16}$ cm$^{-2}$.
Kaczmarczyk (2000b) found (1.0~$\pm$ 0.2)~$\times$ 10$^{16}$ cm$^{-2}$ from an analysis
of permitted {\it A$-$X} bands; the agreement between the two determinations
is at the 2-$\sigma$ level. Recall, however, that Kaczmarczyk's model
has fewer cloud components and larger $b$-values, readily accounting for its reduced column
density. 

\section{BAND OSCILLATOR STRENGTHS}

The crossings of vibrational levels in a triplet state and vibrational levels in the $A~^1\Pi$
state occur near certain rotational $J^\prime$ levels. Therefore, mixing coefficients depend
on $J^\prime$ and thus intersystem $f_{\rm band}$-values generated by interaction with $A~^1\Pi$
have a dependence on $T_{\rm ex}$. Whereas all previous calculations of mixing
coefficients focused on the vibrational level of the A state ($v_A$) closest to the perturbing
triplet level, i.e., a single-$v_A$ treatment, Rostas et al. were the first to use multi-$v_A$
computations that included all $A$ levels with $v^\prime$~= 0 to 12.
They derived $f_{\rm band}$-values that for some intersystem bands were in better agreement
with their own experimental results ($T_{\rm ex}$~= 300~K),
and/or with the IS results of F94 ($T_{\rm ex}$~= 4.2~K). Still, the calculated
$f$-values of Rostas et al. are ``a few percent'' off, because the changes in
$f_{v^\prime v^\prime}$$_{^\prime}$ of the ``parent'' {\it A$-$X} bands
induced by mixing with other triplet levels were ignored. In addition, differences
of a few percent are expected due to the adopted sources for {\it A$-$X} $f$-values: MN and
we used values from \citet{chan}, while Rostas et al.'s preference
lies with values published by \citet{e99}.

The MN line oscillator strength, $f_{J^\prime J^\prime}$$_{^\prime}$,
which we use as input for the fits, includes $f_{v^\prime v^\prime}$$_{^\prime}$
for the permitted {\it A$-$X} band and the usual H$\ddot{\rm o}$nl-London rotational factor,
as well as the single-$v_A$ strength of the perturbation. According to Rostas et al.,
for triplet levels that cross the $A$ levels at low values of $J^\prime$,
single-$v_A$ and multi-$v_A$ calculations should not differ by a large factor
since the interaction with the nearest ``parent'' $v_A$ level is predominant.
Therefore, using $f_{J^\prime J^\prime}$$_{^\prime}$-values from MN is a valid
approach while searching for relatively small corrections due to multi-$v_A$ interactions.
This condition is appropriate for all bands studied here, except for $d$7, $d$10,
and to a lesser extent, $e$4. On the other hand, intersystem bands that are farther
from their nearest parents have ``$J$-independent'' $f_{\rm band}$-values due to their
high-$J^\prime$ crossing, especially under IS conditions.
Again, using $f_{J^\prime J^\prime}$$_{^\prime}$-values from MN poses no difficulties
for scaling the integrated band $f$-values thanks to negligible high-$J^\prime$$^\prime$ level
populations. Indeed, we made no attempt to fit rotational lines independently
of each other; our syntheses did not indicate a need to adjust
individual $f_{J^\prime J^\prime}$$_{^\prime}$-values. Rather than varying the $f$-values, our
variable parameters were $N$($^{12}$CO), which scales the entire band, and $T_{\rm ex}$, which
determines its shape. Eleven $^{12}$CO bands exhibit similar excitation temperatures with
averages of $T_{1,0}$~= 6.1~$\pm$ 1.0~K and $T_{2,0}$~= 5.9~$\pm$ 0.4~K.
Lower $T_{\rm ex}$ were suggested by the fit to $a^\prime$14,
presumably due to effects of higher $\tau$. Although
$f_{\rm band}$ depends on $T_{\rm ex}$, the difference in $f$-values between
$T_{\rm ex}$~= 4.2~K and $T_{\rm ex}$~= 6.0~K is $\la$~1\%, which we ignore here. 
One additional variable fitting parameter was the heliocentric radial velocity. For the main
component, the average value from nine $^{12}$CO intersystem bands is 14.7~$\pm$ 1.0~km~s$^{-1}$,
where the 1~$\sigma$ error is dominated by STIS wavelength calibration uncertainties.
This radial velocity is in very good agreement with the \ion{K}{1} fits mentioned in \S~2.

\subsection{Bands Previously Seen Toward $\zeta$ Oph}

Table 1 lists results of synthetic fits for 12 intersystem bands detected in our spectrum of X Per.
For the reference $f$-values we used the MN $f_{J^\prime J^\prime}$$_{^\prime}$-values
of the $R$(0) lines, which agree to $\pm$~1\% of $f_{\rm band}$-values calculated
for $T_{\rm ex}$~= 4.2~K, except for $a^\prime$14 and $a^\prime$17, which have low-$J^\prime$
crossings and for which the $R$(0) $f$-values are off by $\la$~2\%.
Each ratio of a fitted $f_{\rm band}$-value over the reference MN value was computed from the
ratio of each band's fitted $N$($^{12}$CO) over the actual $N$ determined in \S~3.
Column 5 lists the resulting band $f$-values toward X Per.
For comparison, in columns 6$-$8 we list ratios of $f_{\rm band}$ from this paper, from F94,
and from theoretical predictions in Rostas et al. over the values from MN.
The Rostas et al. values are based on multi-$v_A$ calculations
while the MN values refer to single-$v_A$ results.
The theoretical ratios in column 8 lack formal error bars, although deviations
due to the presence of other triplet levels and to absolute scales should amount to a few
percent. The last column directly compares $f$-values from X Per to those from Rostas et al.

The first seven bands listed in Table 1 were previously detected toward $\zeta$ Oph by F94.
As discovered by F94, the measured $f$-value of $d$7 toward X Per is five times greater than
the value listed by MN in their Tables 5 and 14. However, MN list another $f$-value for $d$7
in their Table 7, based on measurements by Eidelsberg et al. (1992);
the latter result is more in line with both IS $f$-values and was
confirmed in room temperatures experiments by Rostas et al. The more rigorous multi-$v_A$
calculation reproduces the astronomical $f$-values much better than a single-$v_A$ prediction
because $d$7 is relatively far from its nearest parent, $A$2. Our new $f$-value is 6\%
(0.4~$\sigma$) lower than the prediction by Rostas et al., while the F94 result differs
by 1\% only. Consequently, we do not confirm the recent laboratory result by Stark et al. (2002)
that $f_{\rm band}$($d$7) should be reduced by (29~$\pm$ 15)\% relative to Rostas et al.

Our synthesis of the optically thick $a^\prime$14 toward X Per returns an $f$-value 6\%
smaller than that listed by MN, $versus$ the F94 result of $-$24\%. The agreement
with the newer predictions is better: the F94 $f$-value is lower by 19\%, whereas the X Per
result is identical to that of Rostas et al. According to them, single-$v_A$ and multi-$v_A$
$f$-values are also identical: 8.3~$\times$ 10$^{-4}$. Thus, ironically, the strong $a^\prime$14
cannot help to distinguish between the two methods because its (single) interaction with $A$4 is
extremely predominant. All other five bands, $a^\prime$11, $a^\prime$17, $d$12, $e$4, and $e$5,
are also found to be in very good agreement with the multi-$v_A$ predictions of
Rostas et al., i.e., to within 20\%. Of these, the F94 $a^\prime$ bands already
showed an excellent agreement with MN ($\leq$ 6\%), and are within 7\% of Rostas et al.
The last three F94 bands showed large differences (down to $-$49\%) from MN predictions.
All three show better agreement with Rostas et al.: $d$12 now differs by $-$16\%, $e$5 differs
by $-$23\%, while $e$4 is off by $-$37\% (3.1~$\sigma$). The F94 $f$-value for $e$4 remains their
most discrepant result (after $d$7). This weakest band of F94 was affected by an improper
continuum placement, forcing its exclusion from our derivation of $N$($^{12}$CO) in \S~3.
Stark et al.'s (2002) result for $e$4, that its $f$-value is marginally too small by
(17~$\pm$ 15)\% relative to Rostas et al., is very similar to the X Per measurement:
($-$11~$\pm$ 15)\%. Out of the seven bands under discussion, the largest X Per difference relative
to Rostas et al.'s predictions belongs to $a^\prime$17 (+20\%~= +1.0~$\sigma$), a band for which
a few weaker $J^\prime$$^\prime$~= 2,3 lines are blended with another intersystem band, $d$12.

\subsection{Bands Newly Detected Toward X Per}

In our spectrum there are first IS detections of four $^{12}$CO intersystem
bands: $d$8, $d$9, $d$10, and $e$12. Of these, $f_{\rm band}$ was predicted
by Rostas et al. for $d$8, while single-$v_A$ $f$-values were listed by MN for $d$8 and $e$12.
We identified $d$9 and $d$10 as intersystem CO features
thanks to their characteristic shapes. Final confirmation was provided by a comparison
with laboratory bandhead wavelengths in Table 27 of \citet{ts72},
or through the use of molecular state constants from Tilford \& Simmons to compute
reasonably accurate line positions. The newly detected bands are shown in Fig. 1.
Our wavelength calculations with the constants for the $d~^3\Delta_1$ state identify the new
{\it d$-$X} bands as the $F_3$ ($\Omega$~= 1) branch of the $d~^3\Delta_i$ state.
According to the lists of MN, $F_2$ ($\Omega$~= 2) components are at least ten times weaker.

Whereas $d$8 was already mentioned by MN, who adopted Eidelsberg et al.'s (1992)
room temperature $f$-value, the newer experiments of Rostas et al. led to a refined value.
Our determination clearly favors the new value of Rostas et al. calculated for an IS temperature,
since the X Per $f$-value is $-$24\% ($-$1.7~$\sigma$) away from their value,
but is $-$61\% ($-$7.6~$\sigma$) away from the value listed by MN.
The strongest of the new bands, $d$9, has a $W_{\lambda}$ between those of
$d$7 and $d$12. Its strong detection contradicts its non-inclusion in the tables of
MN, who imposed a limiting $f$-value smaller than inferred here. The third {\it d$-$X} band,
$d$10, has the smallest secure $f$-value found in this study, $\sim$~6~$\times$ 10$^{-6}$.
The $d$10 band also lacks a previously published $f$-value. Toward X Per, it is blended
with the red wing of the much wider \ion{Si}{4} $\lambda$1402, which was successfully removed
by profile rectification. The new IS detection of $e$12
close to $A$8 confirms to within 7\% the $f$-value given by MN,
but it was not included in the study of Rostas et al. However, in the last column
of Table 1 we compare $d$9, $d$10, and $e$12 $f$-values with unpublished multi-$v_A$ calculations
(M. Eidelsberg, private communication). Two bands agree to $\pm$ 12\% with these predictions,
while the $f$-value of our faintest band, $d$10, is $-$34\% ($-$2.3~$\sigma$) away from its
predicted value.
 
The high $N$(CO) toward X Per also provides us with a clean detection of a fully resolved
intersystem band from the $^{13}$CO isotopomer, namely, $d$12. This band was
suggested to exist by MN in the $\zeta$ Oph spectrum of $A$6
analyzed by \citet{shef}. For $^{13}$CO, the $d$12 band was predicted to have substantial mixing,
borrowing 19\% of the $A$6 $f$-value according to MN, versus the mere 1\% borrowed by $d$12 from
$A$6 in $^{12}$CO. Since the $A$8 band of $^{13}$CO has $\tau~<$ 1, we could find $N$($^{13}$CO)
toward X Per directly. Using the four-component model from $^{12}$CO, we derived $N$($^{13}$CO)~=
(1.94~$\pm$ 0.08)~$\times$ 10$^{14}$ cm$^{-2}$, which yields a $^{12}$CO/$^{13}$CO
ratio of 73~$\pm$ 12 toward X Per. As can be seen in Table 1, the measured $f$-value of
$^{13}$CO $d$12 is 13\% (2~$\sigma$) larger than the single-$v_A$ prediction in MN.
Unfortunately, multi-$v_A$ calculations for $^{13}$CO have yet to be performed.

\section{CONCLUDING REMARKS}

The agreement between the band $f$-values determined from our measurements of absorption
along the IS line of sight to X Per and the calculations of Rostas et al. is very good.
Rostas et al. suggested that multi-$v_A$ computations are superior for intersystem $f$-values
whenever the spin-forbidden band is not fully overlapped by the permitted band. We concur,
because the 11-band average of column 9 in Table 1 is 0.95~$\pm$ 0.16, i.e., the $-$5\% difference
can be accounted for by known $T_{\rm ex}$ approximations and {\it A$-$X} $f$-value uncertainties,
the latter estimated to be $\approx$~10\% by Rostas et al. As a caveat, had we retained
the $e$4 band in the calculation of $N$($^{12}$CO) in \S~3, the average
would have changed to 0.90~$\pm$ 0.15 due to a systematic adjustment of all $f$-values by $-$5\%.
Individually, nine intersystem bands show very good agreement with their predictions: $a^\prime$11,
$a^\prime$14, $d$7, and $d$12 differ by $\leq~0.4~\sigma$ (7\%), while $a^\prime$17, $e$4, $e$5,
$d$9, and $e$12 agree to within 1~$\sigma$ (20\%). Such small differences between empirical
and predicted $f$-values may be primarily attributed to
assigned observational uncertainties ($\leq$ 17\% for the STIS data; 16\% for $N$ toward X Per)
rather than to the possibility that current quantum-mechanical treatments are inadequate.
The bands $d$9, $d$10, and $e$12 were not included in Rostas et al., but a comparison with
unpublished multi-$v_A$ calculations shows that two $f$-values agree to within 12\%, while
$d$10, our weakest and deblended band, is $-$34\% off. The agreement between empirical results and
theoretical predictions assures that accurate column densities can be extracted from IS spectra
showing very strong absorption in the permitted bands of $^{12}$CO. Finally, there is a need
for theoretical multi-$v_A$ calculations of triplet-singlet mixing coefficients in $^{13}$CO. 

\acknowledgments
We thank Dr. M. Eidelsberg and Dr. D. E. Welty for providing results before publication,
and an anonymous referee for helpful comments.
The research presented here was supported in part by a NASA grant for $HST$ program
GO-08622 and NASA grant NAG5-11440 to U. Toledo.

\begin{deluxetable}{llrcccccccc}
\tablewidth{0pt}
\tabletypesize{\scriptsize}
\tablecaption{Intersystem Bands of CO Toward X Per}
\tablehead{
\colhead{Isotope}
&\colhead{Band}
&\colhead{$W_\lambda$ (m\AA)}
&\colhead{$W_\lambda^{\rm XP}$/$W_\lambda^{\rm \zeta Oph}$}
&\colhead{$f$(X Per)}
&\colhead{$f_{\rm XP}$/$f_{\rm MN}$}
&\colhead{$f_{\rm F94}$/$f_{\rm MN}$}
&\colhead{$f_{\rm R00}$/$f_{\rm MN}$}
&\colhead{$f_{\rm XP}$/$f_{\rm R00}$}
}
\startdata
$^{12}$CO & $a^\prime$11 [$A$2]& 6.52$\pm$0.39 & 4.6$\pm$0.8 & 3.04$\pm$0.51($-$5) & 1.06$\pm$0.18 & 1.06$\pm$0.18 & 1.14 & 0.93$\pm$0.16\\
& $a^\prime$14 [$A$4]&44.67$\pm$0.40 & 2.6$\pm$0.1 & 8.31$\pm$1.30($-$4) & 0.94$\pm$0.15 & 0.76$\pm$0.04 & 0.94 & 1.00$\pm$0.16\\
& $a^\prime$17 [$A$6]& 6.38$\pm$0.29 & 6.0$\pm$1.2 & 3.03$\pm$0.49($-$5) & 1.19$\pm$0.19 & 1.01$\pm$0.20 & 1.00 & 1.20$\pm$0.20\\
& $d$7 [$A$2]        & 4.73$\pm$0.21 & 5.0$\pm$0.7 & 1.92$\pm$0.31($-$5) & 4.78$\pm$0.77 & 5.00$\pm$0.75 & 5.07 & 0.94$\pm$0.15\\
& $d$12 [$A$6]       & 9.62$\pm$0.23 & 5.3$\pm$0.7 & 5.26$\pm$0.83($-$5) & 0.67$\pm$0.11 & 0.58$\pm$0.07 & 0.69 & 0.97$\pm$0.15\\
& $e$4 [$A$2]        & 4.33$\pm$0.33 & 7.2$\pm$1.6 & 1.74$\pm$0.30($-$5) & 0.72$\pm$0.13 & 0.51$\pm$0.10 & 0.81 & 0.89$\pm$0.15\\
& $e$5 [$A$3]        &15.99$\pm$0.23 & 6.8$\pm$0.5 & 7.70$\pm$1.21($-$5) & 0.78$\pm$0.12 & 0.53$\pm$0.04 & 0.69 & 1.13$\pm$0.18\\
& $d$8 [$A$3]        & 2.09$\pm$0.27 & \nodata     & 8.63$\pm$1.75($-$6) & 0.39$\pm$0.08 & \nodata      & 0.52   & 0.76$\pm$0.15\\
& $d$9 [$A$4]        & 6.89$\pm$0.26 & \nodata     & 3.42$\pm$0.55($-$5) & \nodata       & \nodata      &\nodata &(0.88$\pm$0.14)\\
& $d$10 [$A$5]       & 1.30$\pm$0.22 & \nodata     & 5.65$\pm$1.30($-$6) & \nodata       & \nodata      &\nodata &(0.66$\pm$0.15)\\
& $e$12 [$A$8]       & 3.09$\pm$0.24 & \nodata     & 1.52$\pm$0.26($-$5) & 1.07$\pm$0.19 & \nodata      &(0.96)  &(1.12$\pm$0.19)\\
$^{13}$CO & $d$12 [$A$6]       & 4.59$\pm$0.19 & \nodata     & 1.73$\pm$0.11($-$3)    & 1.13$\pm$0.07 & \nodata      &\nodata & \nodata\\
\enddata
\tablecomments{The nearest ``parent'' {\it A$-$X} permitted band is given in square brackets.
References X Per (or XP), MN, F94, and R00 are this paper, Morton \& Noreau (1994), Federman et al.
(1994), and Rostas et al. (2000). The $f$-values are computed for $N$($^{12}$CO)~=
1.41~$\times$ 10$^{16}$ cm$^{-2}$, and for $N$($^{13}$CO)~= 1.94~$\times$ 10$^{14}$ cm$^{-2}$.
Since R00 did not have $f$-values for $d$9, $d$10, and $e$12, the comparison is with unpublished
results from M. Eidelsberg.}
\end{deluxetable}

\begin{figure}
\plotone{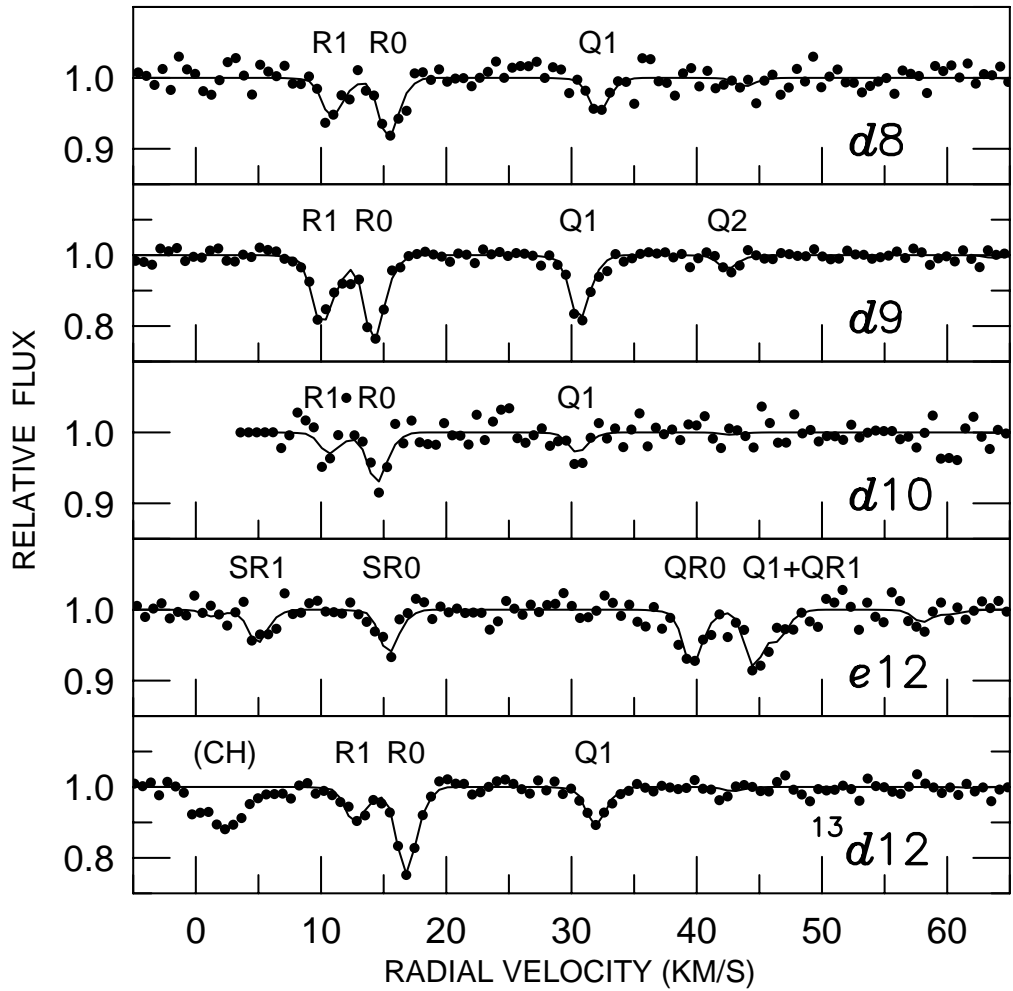}
\caption{Montage of five newly detected intersystem CO bands toward X Per.
STIS data are shown as points, while a solid line depicts individual fits for the 4-cloud
model of $^{12}$CO. Heliocentric radial velocity is shown for the $R$(0) line of all bands.}
\end{figure}

\end{document}